\documentclass[prd, preprint, nofootinbib]{revtex4}

\usepackage{graphicx}
\usepackage{amsfonts}
\usepackage{latexsym}
\usepackage{amsmath}
\usepackage{amssymb}
\usepackage{epsfig}

\begin{document}

\title{ Gravitational waves from breaking of an extra $U(1)$\\ in $SO(10)$ grand unification}

\author{Nobuchika Okada}
 \email{okadan@ua.edu}
 \affiliation{
Department of Physics and Astronomy, 
University of Alabama, Tuscaloosa, Alabama 35487, USA}

\author{Osamu Seto}
 \email{seto@particle.sci.hokudai.ac.jp}
 \affiliation{Institute for the Advancement of Higher Education, Hokkaido University, Sapporo 060-0817, Japan}
 \affiliation{Department of Physics, Hokkaido University, Sapporo 060-0810, Japan}

\author{Hikaru Uchida}
 \email{h-uchida@particle.sci.hokudai.ac.jp}
 \affiliation{Department of Physics, Hokkaido University, Sapporo 060-0810, Japan}

%\date{\today}
%

%%%%%%%%%%%%%%%%%%%%%%
\begin{abstract}
%%%%%%%%%%%%%%%%%%%%%%
In a class of gauged $U(1)$ extended Standard Models (SMs),  
  the breaking of the $U(1)$ symmetry is not only a source for Majorana masses of right-handed (RH) neutrinos 
  crucial for the seesaw mechanism, but also a source of stochastic gravitational wave (GW) background. 
Such $U(1)$ extended models are well-motivated from the viewpoint of grand unification.
In this paper, we discuss a successful ultraviolet completion of a $U(1)$ extended SM 
  by an $SO(10)$ grand unified model through an intermediate step of $SU(5) \times U(1)$ unification.
With a parameter set that is compatible with the $SO(10)$ grand unification, 
  we find that a first-order phase transition associated with the $U(1)$ symmetry breaking 
  can be strong enough to generate GWs with a detectable size of amplitude.   
We also find that the resultant GW amplitude reduces and its peak frequency becomes higher 
  as the RH neutrino masses increase.
%%%%%%%%%%%%%%%%%%%%%%
\end{abstract}
%%%%%%%%%%%%%%%%%%%%%%

%\pacs{}

\preprint{EPHOU-20-006} 

\vspace*{1cm}
\maketitle

%==================================%
%          Main body               %
%==================================%

%%%%%%%%%%%%%%%%%%%%%%%
\section{Introduction}
%%%%%%%%%%%%%%%%%%%%%%%

An extra $U(1)$ gauge interaction is one of the promising and interesting extensions of the standard model (SM)
 of particle physics.
Since tiny but non-vanishing neutrino masses are a clear evidence for the existence of the beyond the SM,
 one of the simplest and the most interesting models is the one based on
 the gauge group $SU(3)_C \times SU(2)_L \times U(1)_Y \times U(1)_{B-L}$~\cite{Pati:1973uk,Davidson:1978pm,Mohapatra:1980qe,Mohapatra:1980},
 where the additional interaction is from the gauged $U(1)_{B-L}$ (baryon number minus lepton number) symmetry.
In the standard $U(1)_{B-L}$ charge assignment,
 three right-handed (RH) neutrinos have to be introduced to fulfill
 the gauge and gravitational anomaly cancellation conditions.
After Majorana masses of RH neutrinos are generated by the spontaneous $U(1)_{B-L}$ gauge symmetry breaking 
 at a high energy scale, the observed tiny neutrino masses are naturally explained 
 by the so-called seesaw mechanism  with the heavy Majorana RH neutrinos
 through their Yukawa interactions with the SM left-handed neutrinos~\cite{SeesawM,SeesawY,SeesawG,SeesawMS}. 
In addition, one of the three RH neutrinos can be a candidate for the dark
 matter in our universe~\cite{Khalil:2008kp,Okada:2010wd,Okada:2016gsh,Okada:2016tci,Okada:2017pgr,Okada:2018ktp}.

Although it is very difficult for any collider experiments to test an additional gauge symmetry if it is broken
 at very high energies, the detection of a gravitational wave (GW) can be a probe for such an extra $U(1)$ symmetry breaking~\cite{Jinno:2015doa,Jinno:2016knw,Chao:2017ilw,Hashino:2018zsi,Okada:2018xdh,Hashino:2018wee,Brdar:2018num,Marzo:2018nov,Hasegawa:2019amx,Haba:2019qol}.
This is because the first-order phase transition in the early universe is
 one of the promising sources of stochastic GW background~\cite{Caprini:2018mtu,Mazumdar:2018dfl,Caprini:2019egz}. 
If a first-order phase transition occurred in the early universe,
 the dynamics of bubble collision~\cite{Turner:1990rc,Kosowsky:1991ua,Kosowsky:1992rz,Turner:1992tz,Kosowsky:1992vn}
 followed by the turbulence of the plasma~\cite{Kamionkowski:1993fg,Kosowsky:2001xp,Dolgov:2002ra,Gogoberidze:2007an,Caprini:2009yp}
 and sonic waves~\cite{Hindmarsh:2013xza,Hindmarsh:2015qta,Hindmarsh:2016lnk} would have generated GWs, 
 which can be detected by the future experiments,  
 such as Big Bang Observer (BBO)~\cite{Harry:2006fi}, DECi-hertz Interferometer Observatory (DECIGO)~\cite{Seto:2001qf},
 Advanced LIGO (aLIGO)~\cite{Harry:2010zz}, and Einstein Telescope (ET)~\cite{Punturo:2010zz}.

In this paper, we consider an ultraviolet (UV) completion of such an extra $U(1)$ extended SM.
A primary candidate scenario for the completion is the Grand Unified Theory (GUT), 
  in which all the SM gauge interactions are unified into a single gauge interaction at a high energy scale.
In this paper, we consider an $SO(10)$ GUT model, in which the extra $U(1)$ gauge group along with the SM gauge group
  is embedded, and  all the SM fermions and RH neutrinos in each generation are also unified
  into a single $\mathbf{16}$ representation of $SO(10)$
 (see, for example, Ref.~\cite{Fukuyama:2012rw} and references therein).
Among several possible paths of symmetry breaking from the $SO(10)$ to the SM gauge group, 
   we consider the following:  
First, $SO(10)$ breaks to $SU(5) \times U(1)$ at a very high scale $M_{\mathrm{SO(10)}}$. 
Next, the $SU(5)$ breaks to the SM gauge group at a scale $ M_{\mathrm{SU(5)}} \simeq 10^{16}$ GeV.
As usual, we simply assume a suitable cosmological inflation scenario by which
 monopoles generated by the $SU(5)$ breaking were diluted away.
The extra $U(1)$ is essentially the gauged $B-L$ symmetry and its breaking
   can take place at any scale below $ M_{\mathrm{SO(10)}}$.
If the extra $U(1)$ symmetry breaking scale is very high, cosmic strings can
   be the dominant source for stochastic GWs~\cite{Dror:2019syi,Buchmuller:2019gfy,Blasi:2020wpy,King:2020hyd}.
Another promising GW source is the first-order phase transition in the early universe
  associated with the $U(1)$ symmetry breaking at a scale 
  lower than about $10^7$ GeV~\cite{Grojean:2006bp,Dev:2016feu,Balazs:2016tbi}.
In previous work on the $U(1)$ extended SMs~\cite{Okada:2018xdh,Hasegawa:2019amx}, 
  we have treated the $U(1)$ gauge coupling as a free parameter and 
  have shown that with its suitable choice the first-order phase transition can generate 
  GWs large enough to be tested in the future experiments. 
However, once we consider the UV completion by the $SO(10)$ GUT, 
   the $U(1)$ gauge coupling is no longer a free parameter and its low energy value 
   is determined by the condition of the gauge coupling unification.  
In this paper, we will examine whether the parameter set compatible with the $SO(10)$ unification 
   can generate a GW spectrum of a detectable size. 
\footnote{Another promising $SO(10)$ breaking path is via the Pati-Salam model~\cite{Pati:1974yy}. 
GWs could be generated by a low-energy Pati-Salam phase transition followed by a late-time inflation
 to dilute monopoles~\cite{Croon:2018kqn}.}

This paper is organized as follows: 
In the next section, we describe the outline for the $SO(10)$ unification of 
   the $U(1)$ extended SM based on the gauge group of $SU(3)_C \times SU(2)_L \times U(1)_Y \times U(1)_X$. 
Towards $SO(10)$ unification, we consider an intermediate path with the $SU(5) \times U(1)_X$ unification
   and show the successful embedding into the $SO(10)$ model with unified gauge couplings.  
In Sect.~\ref{sec:u1}, we describe the system of the extra $U(1)_X$ breaking 
   and discuss the first-order phase transition in the early universe 
   by employing the finite-temperature effective potential of the $U(1)_X$ Higgs field.  
In Sect.~\ref{sec:gw}, we introduce the formulas that we adopt to compute 
   the GW spectrum generated by the first-order phase transition and present the resultant GW spectrum
   for various sets of the model parameters. 
We also discuss the model-parameter dependence of the GW spectrum.
The last section is devoted to our summary.

%%%%%%%%%%%%%%%%%%%%%%%
\section{UV completion by SO(10)}
\label{sec:so10}
%%%%%%%%%%%%%%%%%%%%%%%
%\subsection{A scenario}
\subsection{$SO(10) \supset SU(5) \times U(1)_X$ embedding}

As previously discussed, we consider the UV completion of the $U(1)$ extended SM 
  by $SO(10)$ GUT via the intermediate step of $SU(5) \times U(1)_X$ unification. 
To realize this partial unification, 
  we generalize $U(1)_{B-L}$ of the minimal $B-L$ model to $U(1)_X$, 
  under which the charge of an SM field is defined as a linear combination 
  of its hyper-charge and $B-L$ charge, $q_X= Y x +Q_{B-L}$, 
  with $x$ being a real constant~\cite{Appelquist:2002mw,Oda:2015gna}. 
The particle content of this model is listed in Table \ref{table1}. 
Except for the introduction of the new parameter $x$, the model properties are quite similar to
 those of the minimal $B-L$ model,\footnote{
See Refs.~\cite{Okada:2017cvy, Oda:2017zul, Okada:2020cue}
for interesting phenomenology in an extreme case (a ``hyper-charge oriented'' case), 
namely, $|x| \gg1$.  
}  
  which is realized as the special case of $x=0$.

%%%%%%%%%%%%%%%%%%%%%%%%%%%%%%%%%%%%%%
\begin{table}[h]
	\centering
	\begin{tabular}{|c|ccc|c|} \hline	
		 & $SU(3)_C$ & $SU(2)_L$ & $U(1)_Y$  & $U(1)_X $ \\ \hline
		$q_L^i$ & $\mathbf{3}$ & $\mathbf{2}$ & $\frac{1}{6}$ & $\frac{1}{6}x+\frac{1}{3}$ \\
		$u_R^i$ & $\mathbf{3}$ & $\mathbf{1}$ & $\frac{2}{3}$ & $\frac{2}{3}x+\frac{1}{3}$ \\
		$d_R^i$ & $\mathbf{3}$ & $\mathbf{1}$ & $-\frac{1}{3}$ & $-\frac{1}{3}x+\frac{1}{3}$ \\ \hline
		$l_L^i$ & $\mathbf{1}$ & $\mathbf{2}$ & $-\frac{1}{2}$ & $-\frac{1}{2}x-1$ \\
		$e_R^i$ & $\mathbf{1}$ & $\mathbf{1}$ & $-1$ & $-x-1$ \\
		$N_R^i$ & $\mathbf{1}$ & $\mathbf{1}$ & $0$ & $-1$ \\ \hline
		$H$ & $\mathbf{1}$ & $\mathbf{2}$ & $-\frac{1}{2}$ & $-\frac{1}{2}x$ \\
		$\Phi_2$ & $\mathbf{1}$ & $\mathbf{1}$ & $0$ & $-2$ \\ \hline
	\end{tabular}
\caption{
The particle content of the minimal $U(1)_X$ model. 
In addition to the SM particle content ($i=1,2,3$), three RH neutrinos  
  $N_R^i$ ($i=1, 2, 3$) and one $U(1)_X$ Higgs field $\Phi_2$ are introduced.   
}
\label{table1}
\end{table}
%%%%%%%%%%%%%%%%%%%%%%%%%%%%%%%%%%%%%%%%%%%%%%%

We now consider the embedding,  
\begin{equation}
SU(5)\times U(1)_X \supset 
SU(3)_C \times SU(2)_L \times U(1)_Y \times U(1)_X. 
\end{equation}
As in the standard $SU(5)$ GUT~\cite{Georgi:1974sy}, 
 the charge conjugation of right-handed down quarks and left-handed leptons are embedded in $\mathbf{5}^*$ representation of $SU(5)$,
 while left-handed quarks, the charge conjugation of right-handed up quarks, and the charge conjugation of right-handed charged leptons are embedded
 in $\mathbf{10}$ representation: 
\begin{align}
& \mathbf{5}^* \supset d_R^{i \, C} \oplus \ell_L^i ,
& \mathbf{10} \supset q_L^i \oplus u_R^{i \, C} \oplus \ell_R^{i \, C} ,
\end{align}
This quark and lepton unification requires the following two conditions, 
\begin{align}
& \frac{1}{3}x-\frac{1}{3}=-\frac{1}{2}x-1 , %\label{charge:5} 
& \frac{1}{6}x+\frac{1}{3}=-\frac{2}{3}x-\frac{1}{3}=x+1, 
\end{align}
which should be satisfied with a unique $x$ value. 
The solution is $x= -4/5$ and hence the $SU(5)$ unification
leads to a quantization of $U(1)_X$ charge~\cite{Okada:2017dqs}. 
As is well known, the $SU(5)$ GUT normalization for the SM $U(1)_Y$ coupling
 and rescaled charges are 
\begin{align}
& g_Y = \sqrt{\frac{3}{5}} g_1, 
& Q_1 = \sqrt{\frac{3}{5}} Q_Y.
\end{align}

The $SU(5) \times U(1)_X$ can be embedded into $SO(10)$.  
In the following, we list the decomposition of several $SO(10)$ multiplets 
  to $SU(5) \times U(1)_X$ ~\cite{Slansky:1981yr}: 
\begin{align*}
 SO(10)  & \supset  SU(5)\times U(1)_X \\
\mathbf{10} & =  \mathbf{5}(-2/5) + \mathbf{5}^*(2/5) , \\ 
\mathbf{16} & =  \mathbf{1}(1) + \mathbf{5}^*(-3/5) + \mathbf{10}(1/5) , \\ 
\mathbf{45} & =  \mathbf{1}(0) + \mathbf{10}(-4/5) +\mathbf{10}^*(4/5) + \mathbf{24}(0) ,\\ 
\mathbf{126} & =  \mathbf{1}(2) + \mathbf{5}^*(2/5) + \mathbf{10}(6/5) + \mathbf{15}^*(-6/5) + \mathbf{45}(-2/5) + \mathbf{50}^*(2/5) .  
%\mathbf{210} & =  \mathbf{1}(0) + \dots .
\end{align*}
The SM fermions and RH neutrinos are embedded in {\bf 16} representation. 
The SM Higgs doublet ($H$) is embedded in {\bf 10} representation,\footnote{
To be precise, for deriving realistic SM fermion mass matrices, 
  the SM Higgs doublet is identified with a linear combination of 
  $SU(2)_L$ doublets in {\bf 10} and {\bf 126} representations. 
See Eq.~(\ref{lag:YukawaSO10}) for the Yukawa coupling in the $SO(10)$ GUT. 
} 
  while the $U(1)_X$ Higgs field ($\Phi_2$) is in {\bf 126} representation. 
Similarly to the embedding of $U(1)_Y$ into $SU(5)$, the $SO(10)$ GUT normalization of $U(1)_X$ is given by
\begin{align}
& g_X = \sqrt{\frac{5}{8}} g_{\chi}, 
& Q_{\chi} = \sqrt{\frac{5}{8}} Q_X.
\end{align}

For simplicity, we assume the $SO(10)$ symmetry breaking to the $U(1)_X$ extended SM
  by non-zero vacuum expectation values (VEVs) of $\langle \mathbf{1}(0) \rangle$ and  $\langle \mathbf{24}(0) \rangle$ 
  in a {\bf 45}-representation Higgs field: 
\begin{align}
SO(10) 
\underset{\langle \mathbf{1}(0) \rangle}{\longrightarrow} 
SU(5)\times U(1)_X 
\underset{\langle \mathbf{24}(0) \rangle}{\longrightarrow} 
 SU(3)_C \times SU(2)_L \times U(1)_Y \times U(1)_X .
\label{SSB} 
\end{align}
The final $U(1)_X$ breaking can be realized by a non-zero VEV of 
  $\Phi_2^{\dagger} = \mathbf{1}(2) \subset \mathbf{126}$ Higgs field.

In the $SO(10)$ GUT, the Yukawa interactions for the SM fermions are given by
\begin{equation}
\mathcal{L}_\mathrm{Yukawa} \supset Y_{10} \mathbf{16}_f \mathbf{16}_f \mathbf{10}_H 
+ Y_{126} \mathbf{16}_f \mathbf{16}_f \mathbf{126}_H^{\dagger}, 
\label{lag:YukawaSO10}
\end{equation}
where $\mathbf{16}_f$ is a fermion multiplet (the generation index is suppressed), 
  and $\mathbf{10}_H$ and $\mathbf{126}_H$ are Higgs fields. 
Referring the above decomposition, one can see that
  the VEV of $\mathbf{1}(2) \subset \mathbf{126}$ Higgs breaks the $U(1)_X$ symmetry
  and generates Majorana masses of RH neutrinos in $\mathbf{16}_f$ 
  through the Yukawa coupling $Y_{126}$ in Eq.~(\ref{lag:YukawaSO10}).
In the SM gauge group decomposition, the Yukawa interactions include the neutrino Dirac Yukawa couplings 
  of $\overline{l_L} H N_R$.

%%%%%%%%%%%%%%%%%%%%%%%%%%%%
\subsection{Gauge coupling unification to $SU(5)$}
%%%%%%%%%%%%%%%%%%%%%%%%%%%%
%%%%%%%%%%%%%%%%%%%%%%%%%%%%%%%%%%%%%%
%
\begin{table}[h]
	\centering
	\begin{tabular}{|c|ccc|c|} \hline	
		 & $SU(3)_C$ & $SU(2)_L$ & $U(1)_Y$  & $U(1)_X$ \\ \hline
		$Q$ & $\mathbf{3}$ & $\mathbf{2}$ & $\frac{1}{6}$ & $\frac{1}{5}$ \\
		$\bar{Q}$ & $\mathbf{3}^{\ast}$ & $\mathbf{2}$ & $-\frac{1}{6}$ & $-\frac{1}{5}$ \\
		$D$ & $\mathbf{3}$ & $\mathbf{1}$ & $-\frac{1}{3}$ & $\frac{3}{5}$ \\
		$\bar{D}$ & $\mathbf{3}^{\ast}$ & $\mathbf{1}$ & $\frac{1}{3}$ & $-\frac{3}{5}$ \\  \hline
	\end{tabular}
\caption{
Representations of  the vector-like quarks. 
}
\label{table2}
\end{table}
%%%%%%%%%%%%%%%%%%%%%%%%%%%%%%%%

In a non-supersymmetric framework, 
  a simple setup to achieve the unification of the three SM gauge couplings
  is to introduce two pairs of vector-like quarks ($Q +\bar{Q}$ and $D +\bar{D}$) 
  with TeV scale masses, $M_Q$ and $M_D$, respectively. 
Their representations are listed in Table \ref{table2}.   
It has been shown in Refs.~\cite{Amaldi:1991zx,Chkareuli:1994ng,Chkareuli:1996gq,Choudhury:2001hs,Morrissey:2003sc,Gogoladze:2010in,Chen:2017rpn} that in the presence of the exotic quarks, the SM gauge couplings are successfully unified 
  at $M_\mathrm{SU(5)} \simeq 10^{16}$ GeV. 
This unification scale corresponds to the proton lifetime of $\tau_p \simeq 10^{38}$ yr,
  which is much longer than the current experimental lower limit of $\tau(p \rightarrow \pi^0 e^+) \simeq 10^{34}$ yr
  reported by the Super-Kamiokande collaboration~\cite{Miura:2016krn}.
The presence of the exotic quarks can also work for stabilizing the SM Higgs potential~\cite{Chen:2017rpn}. 
There are Yukawa interactions between $Q$ and $D$, and the SM Higgs doublet.
$Q$ and $D$ decay into the SH Higgs boson and SM quarks 
quickly through a small Yukawa coupling constant.

In the $SU(5)\times U(1)_X$ unification, $D +\bar{D}$ are embedded in $(\mathbf{5},3/5)+(\mathbf{5}^*, -3/5)$, 
  which are then embedded in $\mathbf{16}^*+\mathbf{16}$ multiplets in the $SO(10)$ GUT. 
Similarly, $Q +\bar{Q}$ are embedded in $(\mathbf{10},1/5)+(\mathbf{10}^*, -1/5)$ 
  and then in $\mathbf{16}+\mathbf{16}^*$ multiplets. 
To realize a mass splitting which makes only $D +\bar{D}$ light 
  among the components in the $\mathbf{16}^*+\mathbf{16}$ multiplets, 
  we consider the following Yukawa coupling and mass terms: 
\begin{eqnarray}
 \mathcal{L}_Y &=&  \mathbf{16}^*  \left( Y \mathbf{45}_H + M \right) \mathbf{16}  \nonumber\\
&\supset& 
\mathbf{1}^*  \left( \sqrt{\frac{5}{8}} \, Y \mathbf{1}_H  + M \right) \mathbf{1} 
+
\mathbf{5}^* \left( - \frac{3}{5}\sqrt{\frac{5}{8}} \, Y \mathbf{1}_H + Y \mathbf{24}_H + M \right)  \mathbf{5} \nonumber \\
&+&  \mathrm{tr}\left[ 
 \mathbf{10}^*  \left( \frac{1}{5}\sqrt{\frac{5}{8}} \, Y \mathbf{1}_H + Y \mathbf{24}_H + M \right)\mathbf{10} \right],      
\end{eqnarray}
where $\mathbf{45}_H$ is the {\bf 45}-representation Higgs field in the $SO(10)$ GUT, 
  and the second and third lines are the expression under $SU(5) \times U(1)_X$. 
The $SO(10)$ symmetry breaking down to $SU(3)_C \times SU(2)_L \times U(1)_Y \times U(1)_X$ (see Eq.~(\ref{SSB})) 
   generates new mass terms, and we have 
\begin{eqnarray}
 \mathcal{L}_Y \to   
 M_{\bf 1} \; \mathbf{1}^* \mathbf{1} 
+
\mathbf{5}^* (Y  \langle \mathbf{24}_H \rangle  + M_{\bf 5}) \mathbf{5} 
      + \mathrm{tr}\left[\mathbf{10}^* 
     \left(Y  \langle \mathbf{24}_H \rangle + M_{\bf 10} \right)\mathbf{10} \right] ,     
\end{eqnarray} 
with
\begin{eqnarray}
M_{\bf 1} &=& \sqrt{\frac{5}{8}} \, Y \langle \mathbf{1}_H \rangle  + M,  \nonumber \\ 
M_{\bf 5} &=& - \frac{3}{5}\sqrt{\frac{5}{8}} \, Y \langle \mathbf{1}_H \rangle  + M,  \nonumber\\
M_{\bf 10} &=& \frac{1}{5}\sqrt{\frac{5}{8}} \, Y \langle \mathbf{1}_H \rangle  + M. 
\end{eqnarray}
We set the parameters $Y \sim 1$, $\langle \mathbf{1}_H \rangle ={\cal O}(M_{\rm SO(10)})$ 
and  $M ={\cal O}(M_{\rm SO(10)})$. 
By tuning them, we can realize $M_{\bf 5}={\cal O}(M_{\rm SU(5)})$ while $M_{\bf 1} \sim M_{\bf 10} = {\cal O}(M_{\rm SO(10)})$.  
Next, we tune $ Y \langle {\bf 24}_H \rangle  \simeq  M_{\bf 5} \, {\rm diag}(-1, -1, -1, 3/2, 3/2)$ so as to make 
  only $D+\bar{D}$ in the ${\bf 5} +{\bf 5}^* $ multiplet light. 
This procedure is analogous to the triplet-doublet splitting of the $\mathbf{5}$-plet Higgs field in the standard $SU(5)$ GUT.
We apply the same procedure to the ${\bf 16}+{\bf 16}^*$ multiplets including $Q+{\bar Q}$
   to leave only them light. 
This is done by tuning to realize $M_{\bf 10}={\cal O}(M_{\rm SU(5)})$ while $M_{\bf 1} \sim M_{\bf 5} = {\cal O}(M_{\rm SO(10)})$.

%%%%%%%%%%%%%%%%%%%%%%%%%%%%%%%%%%%%%%
%\begin{table}[htb]
%\centering
%\begin{tabular}{|c|c|c|c|c|} \hline	
%	 & $SU(5)$ &  $U(1)_X $ &  $SO(10)$ \\ \hline
%	$d^c_R{}^i, l_L{}^i$ & $\mathbf{5}^*$ & $-\frac{3}{5}$  & \\ %\hline
%	$q_L{}^i, u^c_R{}^i, e^c_R{}^i $ & $\mathbf{10}$ &  $\frac{1}{5}$ &  $\mathbf{16}$\\ %\hline
%	$N^c_R{}^i$ & $\mathbf{1}$ &  $1$  & \\ \hline
%	$\supset Q$ & $\mathbf{10}$ &  $\frac{1}{5}$ &  \\
%	$\supset \bar{Q}$ & $\mathbf{10}^*$ &  $-\frac{1}{5}$ &  $\mathbf{16}^*$ \\
%	$\supset D$ & $\mathbf{5}$ & $\frac{3}{5}$ & $\mathbf{16}$ \\
%	$\supset  \bar{D}$ & $\mathbf{5}^*$ & $-\frac{3}{5}$  &   \\  \hline
%	$\supset H$ & $\mathbf{5}$ &  $ -\frac{2}{5}$  & $\mathbf{10}$ \\ 
%	            & $\mathbf{5}^*$ &  $ \frac{2}{5}$  &         \\ \hline
%	$\Phi_2^{\dagger}$ & $\mathbf{1}$ & $2$  & $\mathbf{126}$\\ \hline
%\end{tabular}
%\caption{
%The particle content and those representations in a $SO(10)$ model.  
%}
%\label{table3}
%\end{table}
%%%%%%%%%%%%%%%%%%%%%%%%%%%%%%%%%%%%%%%%%%%%%%%

Let us now discuss the gauge coupling unification. 
At the one-loop level, the renormalization group (RG) equations of the SM gauge couplings and $U(1)_\chi$ 
  gauge couplings  are given by 
\begin{eqnarray}
 \mu \frac{d \alpha_i^{-1}}{d \mu} = - \frac{b_i}{2 \pi}, 
\end{eqnarray}
where $\alpha_i= g_i^2/4 \pi$ $(i=$1, 2, 3, and $\chi$). 
The beta function coefficients are expressed as
\begin{align}
b_i=-\frac{11}{3}C_2(G)+\frac{2}{3}T(R_f)N_{R_f}+\frac{1}{6}T(R_s)N_{R_s},
\end{align}
 where $C_2(G)$ is the casimir operator of the group $G$, $T(R_{f(s)})$ is the trace of the product of generators
  ($\mathrm{tr}[t^a t^b]=T(R_{f(s)})\delta^{ab}$), and $N_{R_f(s)}$ is the number of fermions (scalars). 
For each gauge coupling constant in the energy range of $M_Q, M_D < \mu < M_{\mathrm{SU(5)}}$, we have 
\begin{align}
SU(3)_C: b_3 =& b_3^{\rm SM} + 
\frac{2}{3} \left ( 2 +1 \right)  = -5 , \label{bC} \\
SU(2)_L: b_2 =& b_2^{\rm SM} + \frac{2}{3} \times 3 = -\frac{7}{6} , \label{bL} \\
U(1)_1: b_1 =& b_1^{\rm SM} + \frac{3}{5} \times \frac{2}{3} 
 \left( \frac{1}{36} \times 6 \times 2 + \frac{1}{9} \times 3 \times 2 \right)= \frac{9}{2} ,\label{bY} \\
U(1)_{\chi}: b_{\chi} =& \frac{5}{8} \times \frac{2}{3} \left\{\left(\frac{1}{25} \times 6 + \frac{1}{25} \times 3 + \frac{9}{25} \times 3 + \frac{9}{25} \times 2 + \frac{1}{25} \times 1 + 1 \right) \times 3 \right. \nonumber \\
& \left. + 2\left(\frac{9}{25} \times 3+\frac{1}{25} \times 6 \right) \right\} 
 +\frac{5}{8} \times  \frac{1}{3} \left(\frac{4}{25} \times 2 + 4 \right) = 6 ,  \label{bX}
\end{align}
%
%\begin{align}
%SU(3)_C: b_3 =& -11+\frac{1}{3}\left((2+1+1) \times 3 + 2(2+1) \right) = -5 , \label{bC} \\
%SU(2)_L: b_2 =& -\frac{22}{3} + \frac{1}{3} \left((3+1) \times 3 + 2 \times 3 \right)+\frac{1}{6} = -\frac{7}{6} , \label{bL} \\
%U(1)_1: b_1 =& \frac{2}{3} \times \frac{3}{5} \left(\left(\frac{1}{36} \times 6 + \frac{4}{9} \times 3 + \frac{1}{9} \times 3 +\frac{1}{4} \times 2 + 1 \right) \times 3 \right. \nonumber \\
%& \left. + 2\left(\frac{1}{9} \times 3 + \frac{1}{36} \times 6 \right)\right) 
%+ \frac{1}{3} \times \frac{3}{5} \left(\frac{1}{4} \times 2 \right) = \frac{9}{2} ,\label{bY} \\
%U(1)_{\chi}: b_{\chi} =& \frac{2}{3} \times \frac{5}{8} \left(\left(\frac{1}{25} \times 6 + \frac{1}{25} \times 3 + \frac{9}{25} \times 3 + 
%\frac{9}{25} \times 2 + \frac{1}{25} \times 1 + 1 \right) \times 3 \right. \nonumber \\
%& \left. + 2\left(\frac{9}{25} \times 3+\frac{1}{25} \times 6 \right) \right) 
% + \frac{1}{3} \times \frac{5}{8} \left(\frac{4}{25} \times 2 + 4 \right) = 6 . \label{bX}
%\end{align}
%
where $b_3^{\rm SM} = -7$, $b_2^{\rm SM}=-19/6$ and $b_1^{\rm SM}=41/10$ 
  are the beta function coefficients from the SM fields.  
In the following analysis, we set the vector-like quark masses to be $M_Q=M_D=1.5$ TeV, 
   which satisfy the latest LHC bounds~\cite{Sirunyan:2018omb,Sirunyan:2019sza}.
One can see that due to the new contributions of the exotic quarks to $b_{2,3}$, 
   the RG evolutions of $g_{2, 3}$ are flattened and hence the two gauge couplings 
   merge at a higher energy scale than that in the SM.

\begin{figure}[t] 
 \centering
    \begin{tabular}{c}
 \begin{minipage}{0.47\hsize}
\centering
\includegraphics[width=7.7cm]{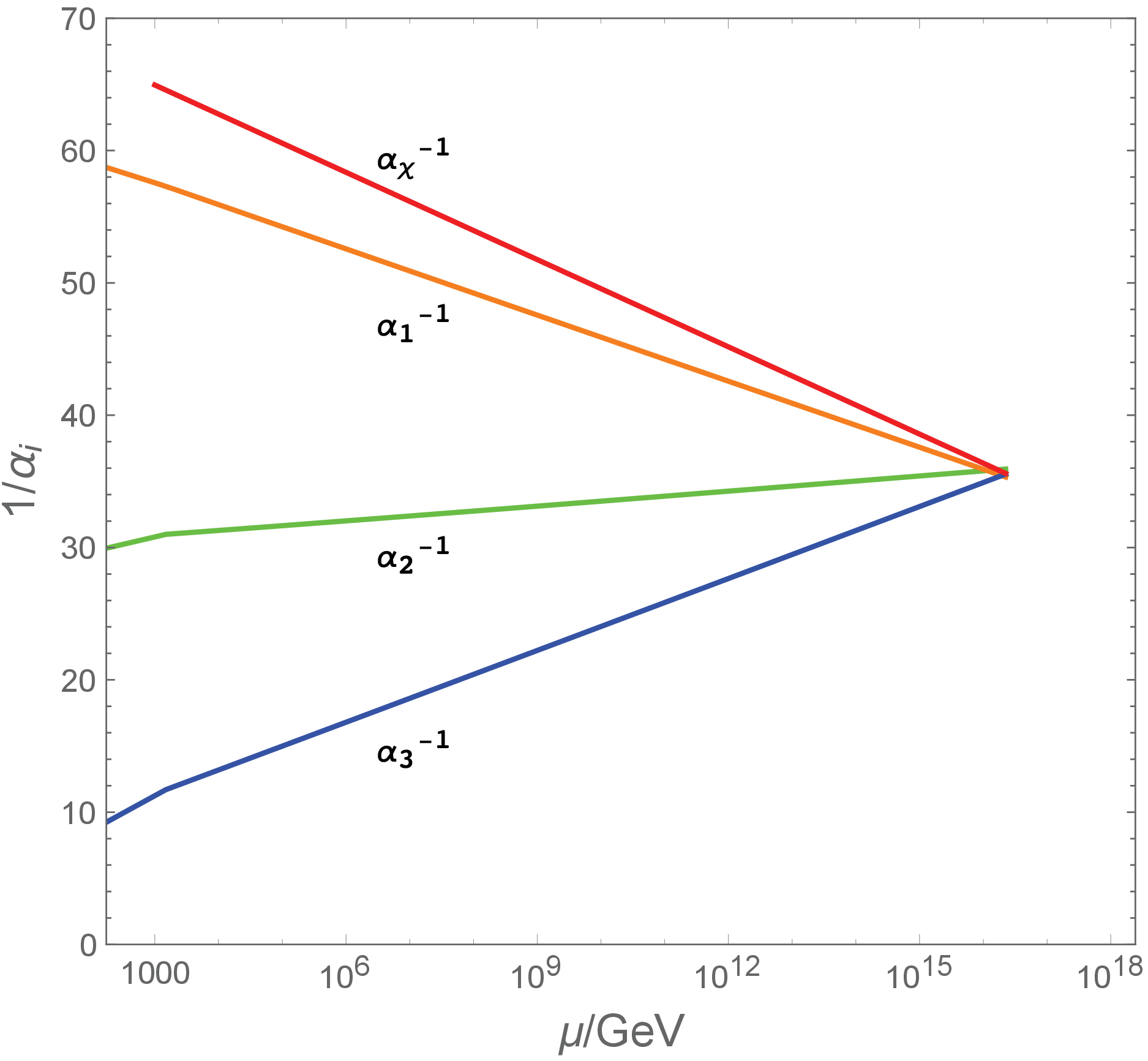}
 \end{minipage}
      \begin{minipage}{0.02\hsize}
        \hspace{2mm}
      \end{minipage}
 \begin{minipage}{0.47\hsize}
\centering
\includegraphics[width=7.7cm]{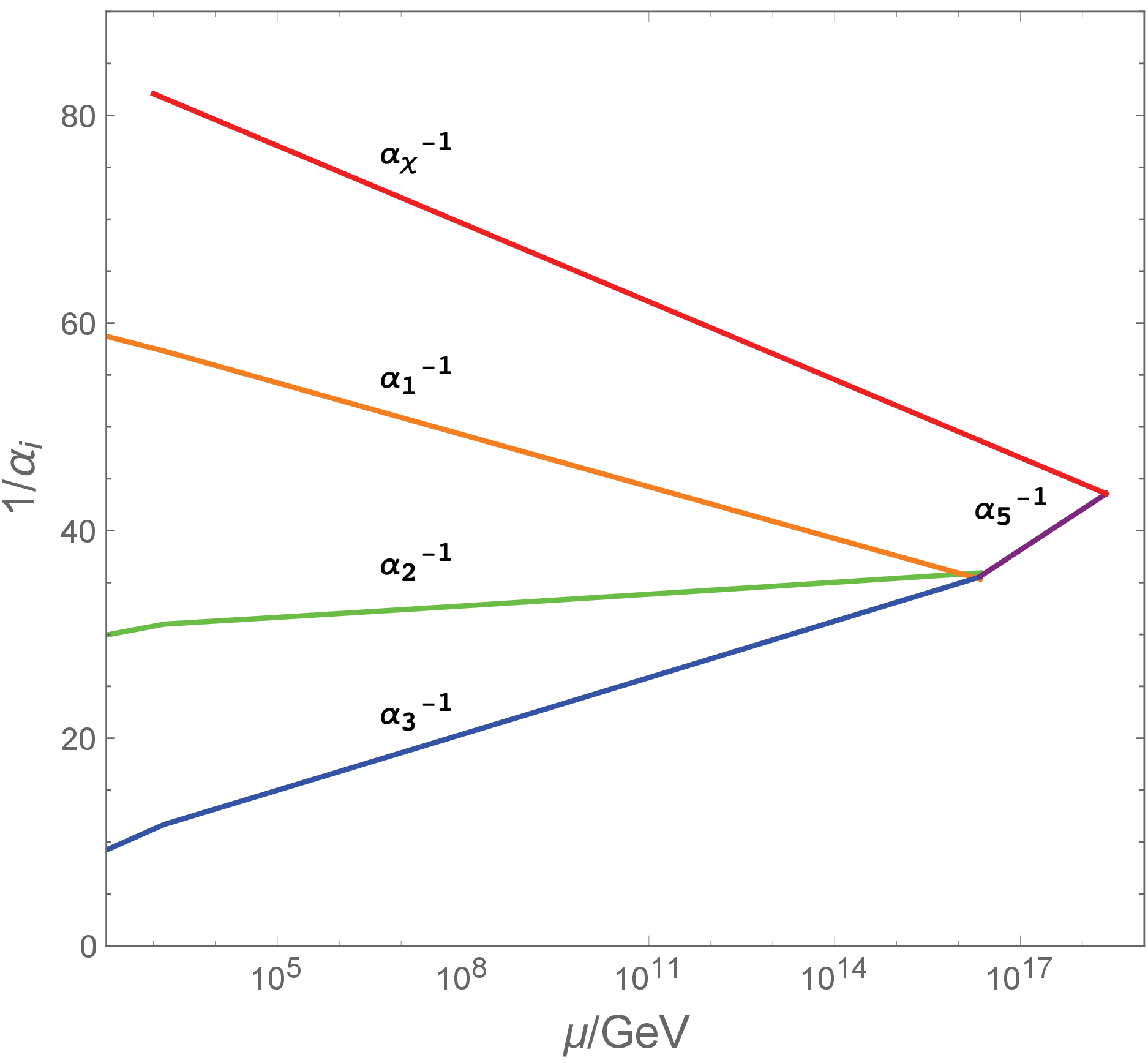}
 \end{minipage}   
 \end{tabular}
\caption{
The RG evolution of the gauge couplings of the $U(1)$ extended SM with the vector-like quarks. 
Three SM gauge couplings are unified at $M_{\rm SU(5)} \simeq 2.24 \times 10^{16}$ GeV. 
{\it Left panel}: The results for the case of $M_{\rm SU(5)}=M_{\rm SO(10)}$. 
{\it Right panel}: The results for $M_{\rm SU(5)} < M_{\rm SO(10)}=M_P$. 
}
\label{Fig:RG}
\end{figure}

In our numerical analysis, we employ the RG equations at the two-loop level 
  (for the beta functions, see, for example, Ref.~\cite{Gogoladze:2010in}) 
  and numerically solve the RG equations with the boundary conditions 
  at the top quark pole mass $\mu=M_t=173.34$ GeV. 
Adopting the fitting formulas given in Ref.~\cite{Buttazzo:2013uya}, we set 
  $g_1(M_t)=0.4626$, $g_3(M_t)=0.6478$, $g_3(M_t)=1.167$, $y_t(M_t)=0.9369$, and 
  $\lambda_H(M_t)=0.2518$, 
  where $y_t$ and $\lambda_H$ are the running top Yukawa and Higgs quartic couplings, respectively. 
We find that three  SM gauge couplings are successfully unified 
  at $M_\mathrm{SU(5)} \simeq 2.24 \times 10^{16}$ GeV. 
Our results are shown in Fig.~\ref{Fig:RG}. 
We will discuss the RG evolution for $g_\chi$ in the next subsection. 

%
%%%%%%%%%%%%%%%%%%%%
%%%%%%%%%%%%%%%%%%%

%%%%%%%%%%%%%%%%%%%%%%%%%%%%
\subsection{Gauge coupling unification to $SO(10)$}
%%%%%%%%%%%%%%%%%%%%%%%%%%%%
After the successful unification of the SM gauge group to $SU(5)$ at $M_{\rm SU(5)}$ 
  we consider the unification of $SU(5) \times U(1)_\chi \to SO(10)$ at $M_{\rm SO(10)}$.  
In the following, let us consider two simple cases: 
  the first is $M_{\rm SU(5)} = M_{\rm SO(10)}$, 
  and the second is $M_{\rm SU(5)} <  M_{\rm SO(10)} =M_P$, 
  where $M_P=2.43 \times 10^{18}$ GeV is the reduced Planck mass. 
The first case is very simple, and the running coupling $g_\chi(\mu$) 
  is determined so as to satisfy the unification condition 
  $g_\chi(M_\mathrm{SU(5)})=g_i (M_\mathrm{SU(5)})$ $(i=1,2,3)$. 
The result is shown in the left-handed panel of Fig.~\ref{Fig:RG}. 
For the second case, we consider the evolution of the $SU(5)$ gauge coupling $g_5$ 
  from $M_{\rm SU(5)}$ to $M_{\rm SO(10)} =M_P$,
  and set the boundary condition for $g_\chi$ as $g_\chi(M_\mathrm{SO(10)})=g_5 (M_\mathrm{SO(10)})$. 

To calculate the RG evolution of $g_5$ and $g_\chi$ in the energy range of 
   $M_{\mathrm{SU(5)}} < \mu < M_{\mathrm{SO(10)}}$,
   we need to know the particle spectrum to determine the beta functions of $g_5$ and $g_\chi$. 
We assume the minimal particle content for the $SU(5) \times U(1)$ theory 
   connecting to the particle contents of the $U(1)_X$ extended SM.    
All the SM fermions and RH neutrinos which are embedded into the three generations
  of {\bf 16}-plets of $SO(10)$ contribute to the beta functions. 
Under the gauge group $SU(5) \times U(1)_X$,     
  the SM Higgs field is embedded into $({\bf 5}, -2/5)$ 
  and the $U(1)_X$ Higgs field is in $({\bf 1}, -2)$. 
In addition, we have an $SU(5)$ adjoint Higgs field with a vanishing $U(1)_X$ charge $({\bf 24}, 0)$, 
  the VEV of which breaks the $SU(5)$ gauge group to the SM gauge group.  
Lastly, as we have discussed in Sect.~\ref{sec:so10} B, 
  the vector-like quarks, ${\bar Q}+Q$ and $D+{\bar D}$, respectively, are embedded
  in the full $SU(5)$ multiplets of ${\bf 10}^* + {\bf 10}$ and ${\bf 5} + {\bf 5}^*$ at $M_{\rm SU(5)}$. 
Taking all these fields into account, the beta function coefficient of the $SU(5)$ gauge coupling 
  at the one-loop level is given by
\begin{align}
b_5 = -\frac{55}{3} +  \frac{2}{3} \left(\frac{1}{2} + \frac{3}{2} \right) \times (3+2)
+ \frac{1}{3} \left( \frac{1}{2} + \frac{1}{2} \times 5 \right) = - \frac{32}{3}.
\label{b5}
\end{align}
The beta function coefficient of the $U(1)_\chi$ gauge coupling 
  at the one-loop level is given by
\begin{align}
b_\chi = \frac{5}{8} \left\{  \frac{2}{3} \left(\frac{9}{25} \times 5  + \frac{1}{25} \times 10 \right) \times (3+2) 
+ \frac{2}{3} \times 3 
+ \frac{1}{3} \left(\frac{4}{25} \times 5  + 4 \right) 
\right \}
 = \frac{41}{6}. 
\label{b5-chi}
\end{align}
With these beta function coefficients and the boundary condition $g_5(M_{\rm SO(10)})=g_\chi(M_{\rm SO(10)})$,  
  we find the solutions for the RG equations. 
Our results are shown in the right-handed panel of Fig.~\ref{Fig:RG}.

%%%%%%%%%%%%%%%%%%%%%%%
\section{Extra $U(1)$ breaking}
\label{sec:u1}
%%%%%%%%%%%%%%%%%%%%%%%
In the low-energy effective theory based on $SU(3)_C\times SU(2)_L\times U(1)_Y\times (1)_X$, 
 the Yukawa interactions of $N_R$ are 
\begin{align}
\mathcal{L}_{Yukawa} \supset  - \sum_{i=1}^{3} \sum_{j=1}^{3} Y^{ij}_{D} \overline{l^i_{L}} H N_R^j 
          -\frac{1}{2} \sum_{k=1}^{3} Y_{N^k} \Phi_2 \overline{N_R^{k~C}} N_R^k 
           + {\rm H.c.} ,
\label{Lag1} 
\end{align}
 where the first term is the neutrino Dirac Yukawa coupling, and the second is the Majorana Yukawa couplings. 
Once the Higgs field $\Phi_2$ develops a nonzero VEV,
 the $U(1)$ gauge symmetry is broken and the Majorana mass terms
 of the RH neutrinos are generated. 
After the electroweak symmetry breaking, tiny neutrino masses are generated 
 through the seesaw mechanism.

In the effective theory, we consider the following tree-level scalar potential: 
\begin{align}
V_0(\Phi_2 )
 = - M^2_{\Phi_2} \Phi_2\Phi_2^{\dagger} +\frac{1}{2}\lambda_2 (\Phi_2\Phi_2^{\dagger} )^2.
\label{potential:tree}
\end{align}
Here, we omit the SM Higgs field ($H$) part and its interaction terms,
 not only for simplicity but also because it has little importance in the following discussion,
 since we are interested in the case that the VEV of the $U(1)_X$ Higgs field is much larger
 than that of the SM Higgs field.

The $U(1)_X$ Higgs field is expanded around its VEV ($v_2$) as
\begin{align}
\Phi_2 =& \frac{ v_2 + \phi_2 + i \chi_2 }{\sqrt{2}} .
\end{align}
The scalar masses are expressed as
\begin{align}
m_{\phi_2}^2 = &  - M^2_{\Phi_2} +\frac{3\lambda_2}{2}v_2^2 , \\
m_{\chi_2}^2 = &  - M^2_{\Phi_2} +\frac{\lambda_2}{2}v_2^2 .
\end{align}
At the classical minimum with $v_2=\sqrt{2 M_{\Phi_2}^2/\lambda_2}$, 
  $\chi_2$ is the would-be Nambu-Goldstone mode eaten by the $U(1)_X$ gauge boson ($Z'$ boson) 
  and $m_{\phi_2}^2=\lambda_2 v_2^2$.
The RH neutrinos $N_R^i$ and the $Z^\prime$ boson acquire their masses as 
\begin{align}
  m_{N_R^i}=& \frac{Y_{N^i}}{\sqrt{2}} v_2, \\ 
  m_{Z^\prime}^2 =& q_{\Phi_2}^2 \, g_X^2 \, v_2^2.
\end{align}

One-loop corrections to the scalar potential for both zero and finite temperatures are
 essential for realizing the first-order phase transition.
One-loop correction is given by
\begin{align}
\Delta V_{1-\mathrm{loop}}(\varphi) = & \sum_s g_s\frac{m_s^4}{64 \pi^2}\left( \ln\frac{m_s^2}{Q^2}-c_s \right)
 - \sum_f g_f\frac{m_f^4}{64 \pi^2}\left( \ln\frac{m_f^2}{Q^2}-c_f \right)  \nonumber \\
  & + \sum_v g_v\frac{m_v^4}{64 \pi^2}\left( \ln\frac{m_v^2}{Q^2}-c_v \right) .
\label{potential:cw} 
\end{align}
Here, $g_i$, with $i=s$ (scalars), $f$ (fermions) and $v$ (vectors) denotes the number
  of internal degrees of freedom, $c_i = 5/6$ $(3/2)$ is a constant for a vector boson
 (a scalar or a fermion),
 and $Q$ is the renormalization scale. 
The finite temperature correction to the effective potential is expressed by
\begin{align}
\Delta V_T(\varphi) =  \sum_s g_s\frac{T^4}{2\pi^2} J_B(m_s^2/T^2)  - \sum_f g_f\frac{T^4}{2\pi^2} J_F(m_f^2/T^2) 
+ \sum_v g_v\frac{T^4}{2\pi^2} J_B(m_v^2/T^2) ,
%\Delta V_T(\varphi) = & \sum_s g_s\frac{T^4}{2\pi^2} J_B(m_s^2/T^2)  - \sum_f g_f\frac{T^4}{2\pi^2} J_F(m_f^2/T^2)  \nonumber \\
%  & + \sum_v g_v\frac{T^4}{2\pi^2} J_B(m_v^2/T^2) ,     
\label{potential:thermal} 
\end{align}
 where $J_{B (F)}$ is an auxiliary function in thermal corrections (see e.g. Refs.~\cite{Bellac,Kapusta}).

We include the thermal correction to masses of $\phi_2$, $\chi_2$ and the $Z'$ boson as given by
\begin{align}
\Delta m^2_{\phi_2 / \chi_2 } =&  \frac{q_{\Phi_2}^2}{4} g_\chi^2 T^2 + \frac{\lambda_2}{6}T^2 + \sum_N\frac{|Y_N|^2}{24}T^2 ,\\
\Delta m^2_{Z'_L} =&  \sum_{\Phi} N_{\Phi}q_{\Phi}^2\frac{g_\chi^2}{6} T^2 + \sum_{f} N_c (q_{Lf}^2+q_{Rf}^2) \frac{g_\chi^2}{6} T^2 ,
\end{align}
 where $q_{\Phi}$ denotes the $U(1)_X$ charge of $\Phi = H$ and $\Phi_2$, respectively, 
  $q_L$ and $q_R$ are those of left- and right-handed fermion $f$,
  $N_\Phi$ is the number of degrees of freedom in $\Phi$, $N_c$ is the color factor, 
  and $\sum_f$ denotes the summation for all fermion flavors.
We have the sum of charges to be
\begin{align}
\sum_{\Phi} N_{\Phi}q_{\Phi}^2=& 4\left(\frac{2}{5}\sqrt{\frac{5}{8}}\right)^2 + 2\left(2\sqrt{\frac{5}{8}}\right)^2 = \frac{27}{5}, \\
\sum_{f} N_c (q_{Lf}^2+q_{Rf}^2) =& 3\left\{ (3+2)\left(-\frac{3}{5}\sqrt{\frac{5}{8}}\right)^2+ (2\times3 +3+1)\left(\frac{1}{5}\sqrt{\frac{5}{8}}\right)^2 + \left(\sqrt{\frac{5}{8}}\right)^2 \right\} \nonumber \\ 
& +3 \left\{ 2\left(\frac{1}{5}\sqrt{\frac{5}{8}}\right)^2+  2\left(-\frac{1}{5}\sqrt{\frac{5}{8}}\right)^2 +\left(\frac{3}{5}\sqrt{\frac{5}{8}}\right)^2 + \left(-\frac{3}{5}\sqrt{\frac{5}{8}}\right)^2  \right\} \label{charge:comp} \\
 = & \frac{153}{20} , 
\end{align}
where the contribution of the second line in Eq.~(\ref{charge:comp}) comes from the vector-like quarks. 
%The thermal masses of the longitudinal mode of the 
%gauge boson as well as the thermal mass of scalars are added and subtracted 
%in the un-resummed Lagrangian, by following to the resummation method~\cite{Carrington:1991hz,Parwani:1991gq,Funakubo:2012qc}.

For our numerical calculations, we have implemented our model
 into the public code \texttt{CosmoTransitions}~\cite{Wainwright:2011kj},
 where both zero- and finite-temperature one-loop effective potentials
 with the corrections for resummation, 
\begin{align}
 & V_{\mathrm{eff}}(\varphi, T) = V_0(\varphi)+ \Delta V_{1-\mathrm{loop}}(\varphi) + \Delta V_T(\varphi, T), 
\label{potential:eff} 
\end{align}
 with $\Phi_2 = \varphi/\sqrt{2}$,
 have been calculated in the $\overline{\textrm{MS}}$ renormalization scheme at a renormalization scale $Q^2=v_2^2$.
Here, as a caveat, we note that there is a long-standing open problem of gauge dependence on the use of the effective Higgs potential.
Our results are also subjects of this issue~\cite{Wainwright:2012zn,Chiang:2017zbz} and should be regarded as a reference value. 

By using \texttt{CosmoTransitions}, we calculate
 the bubble nucleation temperature $T_\star$~\cite{Huber:2008hg}, 
 the latent heat energy density given by
\begin{equation}
\epsilon = \left.\left(V - T\frac{\partial V}{\partial T}\right)\right|_{\{\phi_\mathrm{high}, T_{\star}\}} - \left.\left(V -T\frac{\partial V}{\partial T}\right)\right|_{\{\phi_\mathrm{low}, T_{\star}\}}, 
\end{equation}
 where $\phi_{\mathrm{high}(\mathrm{low})}$ denotes the field value of $\phi$ at the high (low) vacuum
 and the three-dimensional Euclidean action for the bounce solution of the scalar field in the effective potential (\ref{potential:eff}).
We introduce the latent heat to radiation energy density ratio defined by~\footnote{During
 the preparation of this manuscript, a paper by Giese et al. \cite{Giese:2020rtr} appeared. They introduced new parameter $\alpha_{\bar{\theta}}$ which enable us to carry out more accurate calculations.}
\begin{equation}
\alpha \equiv \frac{\epsilon}{\rho_\mathrm{rad}} .
\end{equation}
The radiation energy density is given by
\begin{equation}
\rho_\mathrm{rad} = \frac{\pi^2 g_*}{30}T^4,
\end{equation}
 with $g_*$ being the total number of relativistic degrees of freedom in the thermal plasma.
The bubble nucleation rate per unit volume at a finite temperature is given by
\begin{equation}
\Gamma(T) =  \Gamma_0 e^{-S(T)} \simeq \Gamma_0 e^{-S^3_E(T)/T} .
\end{equation}
Here, $\Gamma_0$ is a coefficient of the order of the transition energy scale,
 $S$ is the action in the four-dimensional Minkowski space, 
 and $S^3_E$ is the three-dimensional Euclidean action~\cite{Turner:1992tz}.
The transition timescale is characterized by a dimensionless parameter
\begin{align}
\frac{\beta}{H_{\star}} \simeq \left. T\frac{d S}{d T}\right|_{T_{\star}} = \left. T\frac{d (S^3_E/T)}{d T} \right|_{T_{\star}}  ,
\end{align}
 with
\begin{equation}
\beta \equiv -\left.\frac{d S}{d t}\right|_{t_\star}. 
\end{equation}
%

%%%%%%%%%%%%%%%%
\section{GW spectrum}
\label{sec:gw}
%%%%%%%%%%%%%%%%
\subsection{GW generation}
%%%%%%%%%%%%%%%%
There are three mechanisms generating GWs by a first-order phase transition:
 bubble collisions, sound waves, and turbulence after bubble collisions.
The resultant spectrum of GW background produced by each of the three mechanisms is expressed as
\begin{equation}
\Omega_{GW}(f) = \Omega_{GW}^\mathrm{coll}(f) +  \Omega_{GW}^\mathrm{sw}(f) +  \Omega_{GW}^\mathrm{turb}(f),
\end{equation}
 in terms of the density parameter $\Omega_{GW}$.
Here, the three terms on the right-hand side denote the GW generated by bubble collisions, sound waves, and turbulence, respectively.

\subsubsection{Bubble collisions}

The GW spectrum generated by bubble collisions for a case of $\beta/H_{\star} \gg 1$
 is fitted with
\begin{align}
\Omega_{GW}^\mathrm{coll}(f)  &= \Omega_{GW}^\mathrm{coll}(f_\mathrm{peak}) \, S^\mathrm{coll}(f), 
\end{align}
with the peak amplitude~\cite{Kosowsky:1992vn}
\begin{align}
h^2 \Omega_{GW}^\mathrm{coll}(f_\mathrm{peak} ) &\simeq 1.7 \times 10^{-5} \kappa_\mathrm{coll}^2\Delta \left(\frac{\beta}{H_{\star}}\right)^{-2}
 \left(\frac{\alpha}{1+\alpha}\right)^2 \left(\frac{g_*}{100}\right)^{-1/3} , %\\
%\Delta &= \frac{0.11 v_b^3}{0.42+v_b^2},%
\end{align}
 the peak frequency 
\begin{align}
f_\mathrm{peak} &\simeq 17 \left(\frac{f_{\star}}{\beta}\right) \left(\frac{\beta}{H_{\star}}\right)
 \left(\frac{T_{\star}}{10^8 \, \mathrm{GeV}}\right)\left(\frac{g_*}{100}\right)^{1/6} \mathrm{Hz} ,\\
\frac{f_{\star}}{\beta} &= \frac{0.62}{1.8-0.1 v_b+v_b^2},
\end{align}
 and the spectral function~\cite{Huber:2008hg}
\begin{align}
S^\mathrm{coll}(f) &= \frac{(a+b)f_\mathrm{peak}^b f^a}{ b f_\mathrm{peak}^{a+b} + a f^{a+b}}, \\
(a, b) &\simeq (2.7, 1.0) .
\end{align}
The efficiency factor for bubble collisions is given by\footnote{An elaborated expression as a function of parameters describing dynamics of bubble walls is given in Ref.~\cite{Ellis:2019oqb}. }
\begin{equation}
\kappa_\mathrm{coll} = \frac{1}{1+ A \alpha}\left( A \alpha +\frac{4}{27}\sqrt{\frac{3\alpha}{2}} \right),
\end{equation}
 with $A = 0.715 $ and 
\begin{align}
\Delta = \frac{0.11 v_b^3}{0.42+v_b^2},
\end{align}
 denotes the bubble wall velocity $v_b$ dependence~\cite{Kamionkowski:1993fg}.

%%%%%%%%%%%%%%%%%
\subsubsection{Sound waves}
%%%%%%%%%%%%%%%%%

The GW spectrum generated by sound waves is fitted by
\begin{align}
\Omega_{GW}^\mathrm{sw}(f)  &= \Omega_{GW}^\mathrm{sw}(f_\mathrm{peak}) \, S^\mathrm{sw}(f), 
\end{align}
with the peak amplitude~\cite{Hindmarsh:2013xza,Hindmarsh:2015qta,Espinosa:2010hh}
\begin{align}
h^2\Omega_{GW}^\mathrm{sw}(f_\mathrm{peak} ) &\simeq 2.7 \times 10^{-6} \kappa_v^2 v_b \left(\frac{\beta}{H_{\star}}\right)^{-1}
 \left(\frac{\alpha}{1+\alpha}\right)^2 \left(\frac{g_*}{100}\right)^{-1/3} (H_{\star} \tau_\mathrm{sw}),
\label{Eq:Amplitude_Sound} % \\
%\kappa_v &\simeq \frac{\alpha}{0.73 + 0.083\sqrt{\alpha}+\alpha },
\end{align}
The efficiency factor ($\kappa_v$) is given by~\cite{Espinosa:2010hh,Ellis:2020nnr}
\begin{equation}
\kappa_v \simeq \left\{
\begin{array}{lll}
  v_b^{6/5} \frac{6.9 \alpha}{1.36-0.037\sqrt{\alpha}+\alpha }  \quad & \textrm{for} & v_b \ll c_s \\
  \frac{\alpha^{2/5}}{0.017+( 0.997+{\alpha})^{2/5} }    & \textrm{for} & v_b \simeq c_s \\
%  \frac{ \sqrt{\alpha} }{0.135+ \sqrt{0.98 + \alpha} }   & \textrm{for} & v_b \simeq v_J \\
  \frac{\alpha_\mathrm{eff}}{\alpha} \frac{\alpha_\mathrm{eff}}{0.73 + 0.083\sqrt{\alpha_\mathrm{eff}}+\alpha_\mathrm{eff} }       & \textrm{for} & v_b \simeq 1 \\
\end{array}
\right. ,
\end{equation}
 with $c_s$ being the sonic speed and $\alpha_\mathrm{eff} = \alpha(1-\kappa_\mathrm{coll})$~\cite{Ellis:2020nnr}. 
In numerical plot shown below, we use the efficiency factor for $v_b \sim 1$,  
 which is realized when friction between the wall and the fluid is weak~\cite{Hindmarsh:2015qta}.
For a given $\alpha$, there is a variation of a factor in this efficiency factor~\cite{Espinosa:2010hh}. 
This may induce a factor or one order of magnitude difference in the final GW spectrum 
 and does not affect our conclusion qualitatively. 
The peak frequency is~\cite{Huber:2008hg}
\begin{align}
f_\mathrm{peak} &\simeq 19 \frac{1}{v_b} \left(\frac{\beta}{H_{\star}}\right)
 \left(\frac{T_{\star}}{10^8 \, \mathrm{GeV}}\right)\left(\frac{g_*}{100}\right)^{1/6} \mathrm{Hz} ,
\label{Eq:Peak_Sound} 
\end{align}
 and the spectral function is~\cite{Caprini:2015zlo}
\begin{align} 
S^\mathrm{sw}(f)= \left(\frac{f}{f_{\mathrm{peak}}} \right)^3 
 \left( \frac{7}{ 4+ 3 \left(\frac{f}{f_\mathrm{peak}}\right)^2 } \right)^{7/2}.
\end{align}
The active period of sound waves is expressed as~\cite{Guo:2020grp,Hindmarsh:2020hop}
\begin{align}
 \tau_\mathrm{sw} &= \frac{1}{H_{\star}} \left( 1- \frac{K^{1/4}}{\sqrt{K^{1/2} +2 H_\star R_\star}} \right) , \\
 K & \simeq \frac{\kappa \alpha}{1+\alpha}, 
\end{align}
 where $R_{\star} \simeq (8\pi)^{1/3} v_b/\beta $ is the average separation between bubbles.
The last factor in Eq.~(\ref{Eq:Amplitude_Sound}) represents the suppression effect due to
 the short-lasting sonic wave as a source of GW generation
 compared with the Hubble time scale ($H_{\star}$) for the case of $H_{\star}\tau_\mathrm{sw} < 1$,
 as pointed out in Refs.~\cite{Ellis:2018mja,Ellis:2019oqb} (see also Refs.~\cite{Cutting:2019zws,Hindmarsh:2019phv,Ellis:2020awk,Schmitz:2020rag}).

\subsubsection{Turbulence}

The GW spectrum generated by turbulence is fitted by
\begin{align}
\Omega_{GW}^\mathrm{turb}(f)  &= \Omega_{GW}^\mathrm{turb}(f_\mathrm{peak}) \, S^\mathrm{turb}(f), 
\end{align}
with the peak amplitude~\cite{Kamionkowski:1993fg}
\begin{align}
h^2\Omega_{GW}^\mathrm{turb}(f_\mathrm{peak} ) &\simeq  3.4 \times 10^{-4} v_b \left(\frac{\beta}{H_{\star}}\right)^{-1}
 \left(\frac{\kappa_\mathrm{turb} \alpha}{1+\alpha}\right)^{3/2}
 \left(\frac{g_*}{100}\right)^{-1/3} ,
\end{align}
 the peak frequency 
\begin{align}
f_\mathrm{peak} &\simeq 27 \frac{1}{v_b} \left(\frac{\beta}{H_{\star}}\right)
 \left(\frac{T_{\star}}{10^8 \, \mathrm{GeV}}\right)\left(\frac{g_*}{100}\right)^{1/6} \mathrm{Hz} , 
\end{align}
 and the spectral function~\cite{Caprini:2009yp,Binetruy:2012ze,Caprini:2015zlo}
\begin{align}
S^\mathrm{turb}(f) &= \frac{\left(\frac{f}{f_{\mathrm{peak}}} \right)^3}{
 (1 + \frac{f}{f_\mathrm{peak}} )^{11/3}(1+\frac{8\pi f}{h_{\star}})  } , \\
h_{\star} &= 17 \left(\frac{T_{\star}}{10^8 \mathrm{GeV}}\right)\left(\frac{g_*}{100}\right)^{1/6} \mathrm{Hz} .
\end{align}
We set the efficiency factor for turbulence to be $\kappa_\mathrm{turb} \simeq 0.05 \kappa_v$.

%%%%%%%%%%%%%%%%%%%%%%%%%%%%%
\subsection{Predicted spectrum for benchmark points}
\begin{figure}[htbp]
%  \begin{center}
\centering
\includegraphics[clip,width=12.0cm]{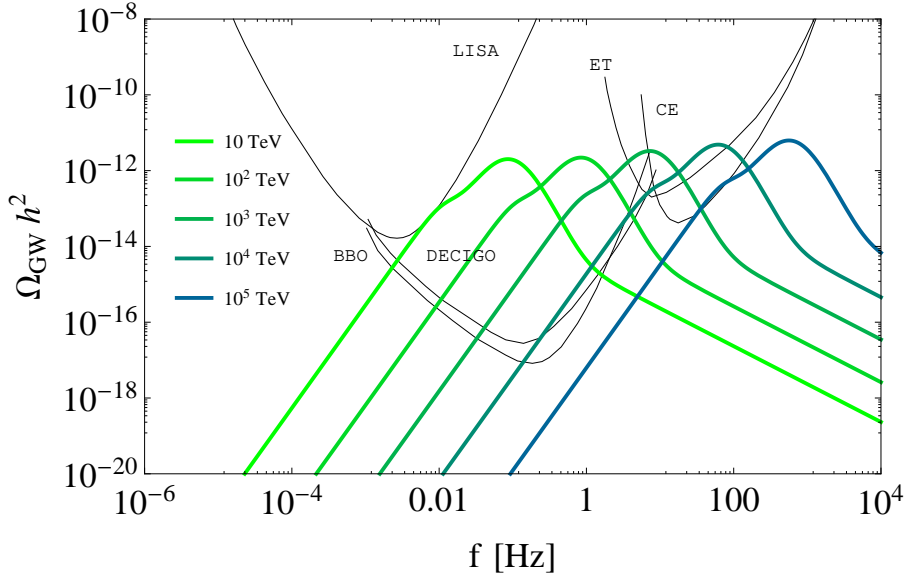}
\caption{
The predicted GW spectrum for various symmetry breaking scales for $\lambda_2=6\times 10^{-4}$. 
The difference of the symmetry breaking scale is indicated by colors as shown in the legends. 
Black solid curves are the expected sensitivities of each indicated experiments derived in Ref.~\cite{Schmitz:2020syl}.
}
\label{Fig:v2_dep}
%  \end{center}
\end{figure}

%%%%%%%%%%%%%%%%%%%%%%%%%%%%%
At first, we show the dependence of the resultant GW spectrum on the energy scale of symmetry breaking, 
  or equivalently, the VEV ($v_2$) scale. 
In Fig.~\ref{Fig:v2_dep} we show the GW spectrum for various symmetry breaking scales 
   for a fixed value of $\lambda_2 = 6\times 10^{-4}$.
As is expected, the peak frequency becomes higher, as symmetry breaking occurs at higher energies. 
The black solid curves denote the expected sensitivities of each indicated experiment: LISA, DECIGO, BBO, ET, and Cosmic Explore (CE). 
The expected sensitivity curves for each experiment are quoted from Ref.~\cite{Schmitz:2020syl}.
We list our results for five benchmark points in Table~\ref{points}.
\begin{table}
\begin{center}
  \begin{tabular}{|cc|ccc|ccc|}
   \hline
 $g_\chi$ &  $Q (=v_2)$ &$\Delta\rho/(0.1 Q)^4$ & $T_\star/(0.1 Q)$ & Action & $\alpha$ & $\beta/H_{\star}$ &\\\hline 
 $0.447$ &  $10$ TeV & $12.8381$ & $0.8865 $ & $125.9341 $ & $0.6318$ & $341.74$ &\\\hline
 $0.455$ &  $10^2$ TeV & $12.8277$ & $0.8728 $ & $116.0031 $ & $ 0.6719$ & $343.69$ &\\\hline
 $0.463$ &  $10^3$ TeV & $12.8105$ & $0.8573 $ & $106.1279 $ & $0.7209$ & $299.61$ &\\\hline
 $0.473$ &  $10^4$ TeV & $12.7969$ & $0.8426 $ & $96.6074 $ & $0.7717$ & $259.75$ &\\\hline
 $0.482$ &  $10^5$ TeV & $12.7719$ & $0.8247 $ & $87.0567 $ & $0.8393$ & $242.64$ &\\\hline
   \end{tabular}
\end{center}
  \caption{
  Input and output parameters for the several benchmark points are listed. 
  In our calculations, we have set $\lambda_2=6\times 10^{-4}$.}
\label{points}
\end{table}

\begin{figure}[htbp]
%  \begin{center}
\centering
\includegraphics[clip,width=12.0cm]{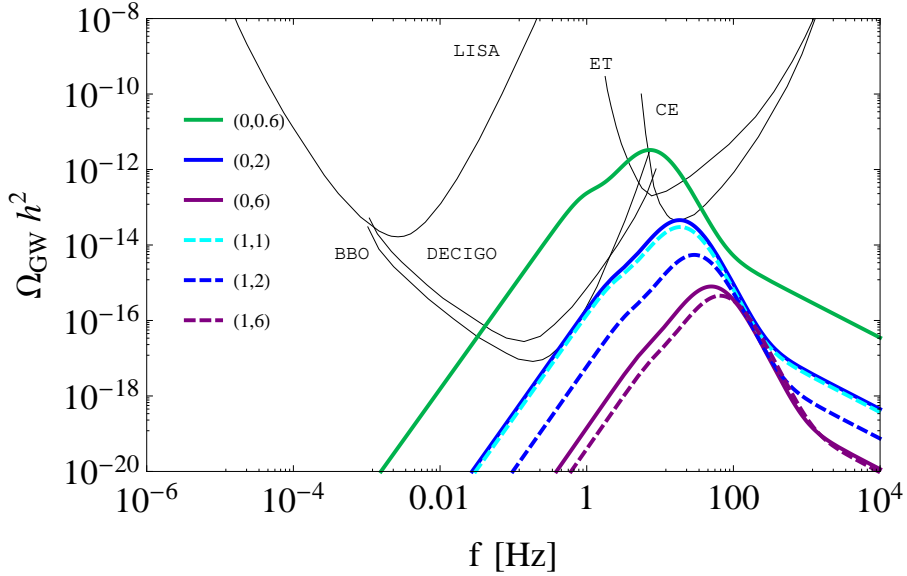}
\caption{
The predicted GW spectrum for various values of $Y_N$ and $\lambda_2$ 
  for $g_\chi=0.463$ and $v_2=1$ PeV. 
Parameters in the legend denote $(Y_N, \lambda_2\times 10^3)$. 
}
\label{Fig:l2Y_dep}
%  \end{center}
\end{figure}

Next, we study how the $U(1)_X$ Higgs quartic coupling and Yukawa coupling affect the resultant GW spectrum.  
In Ref.~\cite{Hasegawa:2019amx}, it has been shown in the context of the minimal $U(1)_{B-L}$ model that 
  the peak amplitude decreases with the peak frequency getting higher as $\lambda_2$ increases, and 
  this dependence is approximated as $\Omega_{\mathrm{GW}}h^2(f_\mathrm{peak}) \propto \lambda_2^{-1/4}$ 
  and $f_{\mathrm{peak}} \propto \lambda_2$.
As $\lambda_2$ increases, the tree level potential (\ref{potential:tree})
 becomes more significant. 
The negative mass squared at the field origin becomes sizable and
 the potential becomes steeper as a whole. 
Then, the strength of the first-order phase transition becomes weaker and $\alpha$ becomes smaller. 
Thus, the amplitude of GW decreases.
We not only reconfirm this $\lambda_2$ dependence in the $SO(10)$ completed model
  but also find that non-vanishing Yukawa coupling has a similar effect on the GW spectrum.
As a Yukawa coupling increases, the peak amplitude decreases with the peak frequency increasing.
This is because fermion loops generate only thermal mass term and not an effective trilinear term 
 in the thermal potential (\ref{potential:thermal}), which weakens the first order phase transition.
Assuming the hierarchy among the Yukawa couplings as $Y_{N_3} \equiv Y_N \gg Y_{N_2}, Y_{N_1}$, 
  for simplicity, we show in Fig.~\ref{Fig:l2Y_dep} the GWs spectrum for various values of 
  $Y_N$ and $\lambda_2$ for $g_\chi=0.463$ and $v_2=1$ PeV
  (see the third benchmark in Table \ref{points}). 
In the figure, we see that two different parameter sets, 
   $( Y_N, \lambda_2) = (0, 0.002)$ and $(Y_N, \lambda_2) = (1, 0.001)$, predict almost the same GW spectrum.  
This is because the dependence of the resultant GW spectrum on $Y_N $ is quite similar to that of $\lambda_2$. 
We have found that for $\lambda_2 \gtrsim 0.006$, 
   the dependence of the GW spectrum on $Y_N$ is weak. 
We list our results for five benchmark points in Table~\ref{c_points}.
\begin{table}
\begin{center}
  \begin{tabular}{|cc|ccc|ccc|}
   \hline
 $Y_N$ &  $\lambda_2$ & $\Delta\rho/(0.1 Q)^4$ & $T_\star/(0.1 Q)$ & Action & $\alpha$ & $\beta/H_{\star}$ &\\\hline 
 $0.$ & $0.002$ & $17.1290$ & $1.3725  $ & $167.3686  $ & $0.1467$ & $477.46$ &\\\hline
 $0.$ & $0.006$ & $33.0336$ & $2.1506  $ & $258.5560 $ & $ 0.0469$ & $821.17$ &\\\hline
 $1.$ & $0.001$ & $2.2088$ & $0.7537  $ & $93.5851 $ & $ 0.2080$ & $892.03$ &\\\hline
 $1.$ & $0.002$ & $4.0694$ & $0.9977  $ & $122.9623 $ & $ 0.1248$ & $1034.35$ &\\\hline
 $1.$ & $0.006$ & $15.2125$ & $1.6639  $ & $201.6852 $ & $ 0.0603$ & $1406.19$ &\\\hline
   \end{tabular}
\end{center}
  \caption{Input and output parameters of several benchmark points varying $\lambda_2$ and $y_N$ are listed.
  We have set $g_\chi=0.463$ and $v_2=1$ PeV in this calculation.}
\label{c_points}
\end{table}
%
%%%%%%%%%%%%%%%%%%%%
%%%%%%%%%%%%%%%%%%%
%
%%%%%%%%%%%%%%%%%%%%
%%%%%%%%%%%%%%%%%%%

%%%%%%%%%%%%%%%%%%%%%%%
\section{Summary}
%%%%%%%%%%%%%%%%%%%%%%%

In this paper, we have considered the $U(1)_X$ extended SM 
   and studied the spectrum of stochastic GWs generated
  by the first-order phase transition associated with the extra $U(1)_X$ symmetry breaking
  in the early universe.
This breaking is responsible for the generation of Majorana masses of RH neutrinos.  
We have investigated a UV completion of the $U(1)_X$ extended SM by an $SO(10)$ GUT.    
In this UV completion, the extra $U(1)$ gauge coupling is unified with the SM gauge couplings,  
  and thus the extra $U(1)$ gauge coupling at the phase transition epoch is no longer a free parameter 
  and $g_\chi \sim 0.4$ from the gauge coupling unification condition.  
We have found that the first-order phase transition triggered by this extra $U(1)$ symmetry breaking
  can be strong enough to generate GWs with a detectable size of amplitude 
  if the $U(1)_X$ Higgs quartic coupling is small enough and the symmetry breaking scale 
  (the bubble nucleation temperature $T_\star$) is smaller than about $10^5$ ($10^4$) TeV.

We have also clarified the dependence of the resultant GW spectrum 
  on the RH neutrino Majorana Yukawa couplings, in other words, the mass scale of RH neutrinos.
As the Yukawa couplings increase, the amplitude of GW background reduces and the peak frequency slightly increases. 
We have found a similar behavior in the GW spectrum as we change the $U(1)_X$ Higgs quartic coupling. 
Thus, different combinations of the Yukawa and the quartic couplings can result in almost the same GW spectrum. 
In order to extract the information about RH neutrino masses from the spectral shape of GW background, 
   the information on the $U(1)_X$ Higgs quartic coupling is necessary.

%======================================%
%<<<<<<<<<< ACKNOWLEDGMENTS >>>>>>>>>>>%
%======================================%

\section*{Acknowledgments}
This work is supported in part by the U.S. DOE Grant No.~DE-SC0012447 (N.O.),
 the Japan Societyfor the Promotion of Science (JSPS) KAKENHI Grants
 No.~19K03860 and No.~19H05091 and No.~19K03865 (O.S.), and 
 the JSPS Research Fellowships for Young Scientists, No.~20J20388 (H.U.).

%\end{document}
 
%======================================%
%<<<<<<<<<<< bibliography >>>>>>>>>>>>>%
%======================================%

%%%%%%%%%%%%%%%%%%%%%%%%%%%%%%%%%%%%%%%%%%%%%%%%%%%%%%%%%%%%

%%%%%%%%%%%%%%%%%%%%%%%%%%%%%%%%%%%%%%%%%%%%%%%%%%%%%%%%%%%%

\end{document}